%% file: main.tex
\documentclass{Interspeech}
\usepackage{amsmath,graphicx}
\usepackage{xcolor,colortbl}
\usepackage{multirow}
\usepackage{array}
\usepackage{cite}
\usepackage{amssymb}
\usepackage{mathrsfs}
\usepackage{blindtext}
\usepackage{algpseudocode}
\usepackage{algorithm}
\usepackage{multirow}
\usepackage{bm}
\usepackage{booktabs}
\usepackage{url}
\usepackage{subcaption}
\usepackage[super]{nth}
\usepackage{float}
\usepackage{arydshln}
\usepackage{amsfonts}
\usepackage{hyperref}
\usepackage{nameref}
\usepackage[skip=5pt]{caption}

\interspeechcameraready

\title{Context-Driven Dynamic Pruning for Large Speech Foundation Models}

\author[affiliation={1}]{Masao}{Someki}
\author[affiliation={1}]{Shikhar}{Bharadwaj}
\author[affiliation={1}]{Atharva Anand}{Joshi}
\author[affiliation={1}]{Chyi-Jiunn}{Lin}
\author[affiliation={1}]{Jinchuan}{Tian}
\author[affiliation={1}]{Jee-weon}{Jung$^{\dagger}$}
\author[affiliation={2}]{Markus}{Müller}
\author[affiliation={2}]{Nathan}{Susanj}
\author[affiliation={2}]{Jing}{Liu}
\author[affiliation={1}]{Shinji}{Watanabe}

\affiliation{Language Technologies Institute}{Carnegie Mellon University}{USA}
\affiliation{Neural Efficiency Science}{Amazon Artificial General Intelligence}{USA}
\email{
msomeki@andrew.cmu.edu}

\usepackage{comment}

\begin{document}

\maketitle

\begin{abstract}
Speech foundation models achieve strong generalization across languages and acoustic conditions, but require significant computational resources for inference.  
In the context of speech foundation models, pruning techniques have been studied that dynamically optimize model structures based on the target audio leveraging external context.  
In this work, we extend this line of research and propose context-driven dynamic pruning, a technique that optimizes the model computation depending on the context between different input frames and additional context during inference.
We employ the Open Whisper-style Speech Model (OWSM) and incorporate speaker embeddings, acoustic event embeddings, and language information as additional context.
By incorporating the speaker embedding, our method achieves a reduction of 56.7 GFLOPs while improving BLEU scores by a relative 25.7\% compared to the fully fine-tuned OWSM model.
\end{abstract}

\renewcommand{\thefootnote}{\fnsymbol{footnote}}
\footnotetext[2]{Currently at Apple.}
\renewcommand{\thefootnote}{\arabic{footnote}}


\input{sections/introduction}

\input{sections/preliminaries_related}


\input{sections/local_gate_predictor}

\input{sections/experiments}

\input{sections/conclusion}

\section{Acknowledgement}
Experiments of this work used the Bridges2 system at PSC and Delta system at NCSA through allocations CIS210014 and IRI120008P from the Advanced Cyberinfrastructure Coordination Ecosystem: Services \& Support (ACCESS) program, supported by National Science Foundation grants \#2138259,\#:2138286, \#:2138307, \#:2137603, and \#:2138296.

\clearpage

\bibliographystyle{IEEEtran}
\bibliography{strings, refs}

\end{document}

%% file: sections/introduction.tex
\section{Introduction}
\label{sec:intro}

Training large neural networks on large-scale speech datasets has achieved significant success in various speech-related tasks~\cite{radford2023robust, nvidia_canary, meta_mms, owsm_v31}.
These speech foundation models demonstrate strong generalization ability and robustness across languages~\cite{meta_mms}, speakers, and acoustic conditions~\cite{radford2023robust}.
While these large models achieve high performance, the use of large-scale models with billions of parameters requires substantial computational resources.
Several approaches have been proposed to mitigate these challenges, including knowledge distillation~\cite{distilwhisper,chang2022distilhubert,colld}, quantization~\cite{4bit_lstm,2bit_conformer,ding2024usm}, pruning~\cite{2102.04010,l0_regularization,lai2021parp,peng2023structured,context_aware}, or combination of these techniques~\cite{dphubert,ding2024usm}.

Among these lines of work, recent research has focused on pruning techniques that dynamically adjust model computation during inference, including
frame skipping or temporal pruning~\cite{tang2023dynamic,zhang19d_interspeech},
early exit~\cite{1603.08983,dan_early_exit,pmlr-v202-xu23g}, 
and structured dynamic pruning for large speech models~\cite{i3d,context_aware}.
Specifically, Someki et al~\cite{context_aware} proposed a dynamic pruning method for speech foundation models that leverages speech features, task characteristics, and language to adapt the model architecture during inference.
They introduced a Global Gate Predictor (globalGP) that generates pruning masks from the input speech and applied dynamic pruning during inference.
However, since this method determines pruning masks for all modules based on input, it cannot account for layer-wise output variations across utterances.

Based on these trends, this study proposes a pruning mechanism named the Local Gate Predictor (localGP), which applies pruning decisions at each layer independently, in contrast to globalGP, which applies a uniform pruning mask across all layers.
We designed localGP to adaptively generate pruning masks based on context that varies across inferences.
We enhance encoder pruning by conditioning on external contexts including speaker embeddings (
ECAPA\_TDNN~\cite{desplanques20_interspeech}),
acoustic event feature (BEATs~\cite{chen2023beats}),
and language information (URIEL lang2vec~\cite{littell2017uriel}), 
We further propose training the localGP at the token level, allowing different modules to be used depending on token history, and analyze the improvements in performance and pruning characteristics.
Since localGP utilizes external context at each layer, it is more effective at incorporating contextual information compared to globalGP, which applies pruning decisions uniformly without considering the internal computations of the model.
Our contributions can be summarized as follows:
\begin{itemize}[noitemsep, topsep=0pt]
    \item We propose localGP, a model designed to utilize various external contexts that dynamically change during inference.
    \item We enhance encoder pruning by integrating speaker embeddings and acoustic events, reducing 56.7 GFLOPs while boosting BLEU by 25.7\% and preserving ASR performance comparable to the fully fine-tuned OWSM model.
    \item Analyzing pruning patterns in Section~\ref{ssec:pruning_pattern_encoder} and Section~\ref{ssec:exp_frame_decoder}, we found that encoder-side pruning behaves similarly to a voice activity detection (VAD) system, while decoder-side pruning shows distinct patterns for specific token types.
\end{itemize}

%% file: sections/preliminaries_related.tex
\section {Preliminaries and Related Works}
\label{sec:global_gp}

\subsection{Related Works}
\label{ssec:related_works}

Pruning techniques for speech processing models have been extensively studied~\cite{rethink_pruning,2102.04010,audio_lottery,peng2023structured}. 
However, in these approaches, the pruning mask is determined either during or after training and remains fixed during inference.
As a result, the sparse pattern is \textit{static} regardless of the input data during inference.
Therefore, while such methods improve computational efficiency, they do not account for contextual information, which may prevent the model from achieving optimal performance.  
To mitigate this limitation, recent research has explored approaches that \textit{dynamically} adjust the model structure during inference~\cite{1603.08983,2207.02393,annrnn,i3d,pmlr-v202-xu23g,context_aware}, enabling more adaptive and context-aware pruning.
Following this trend, our study further advances pruning methods that dynamically modify the model during inference and explores their applicability.
Furthermore, we propose localGP that leverages pretrained models to extract context information, enabling dynamic pruning conditioned on a more diverse set of contextual factors during inference.

Previous research has explored pruning  input data for models~\cite{lu15e_interspeech}.
This approach has been adopted in many speech processing models to shorten input length.
In recent years, dynamic pruning techniques have also been developed in this domain~\cite{zhang19d_interspeech,zhang-etal-2020-adaptive}.
However, these methods either provide frames reduced by a fixed ratio regardless of speech information or perform frame pruning using only speech input.
In this study, we leverage external context captured by localGP to apply optimal input data pruning to each module within the model (e.g., self-attention).
We refer to this approach as \textit{temporal pruning}, as it dynamically subsamples frames based on contextual information rather than pruning entire module computations for all frames in an utterance. 
Additionally, we compare its performance with \textit{utterance-wise pruning}, which removes computations at the utterance level, and discuss the results in Section~\ref{ssec:frame_vs_utt}.


\input{blocks/algorithm1}

\subsection{Dynamic Pruning}
\label{para:dynamic_pruning}

Let $z$ be the pruning mask, $\theta$ the model parameters, $f(\cdot)$ the neural network subject to pruning, and $x$ the input to the model.
The output $\tilde{y}$ of the pruned model is computed as
$\tilde{y} = f(x; \theta \odot z), \quad z \in \{0, 1\}^{|\theta|}$,
where $|\theta|$ represents the number of parameters, and $\odot$ denotes element-wise multiplication.
Typically, $z$ is generated as a binary mask by applying a threshold to probabilities computed by another neural network $g$.
For context-aware dynamic pruning, function $g$ takes the context $C$ as its input.
However, performing such discretization during training results in non-differentiability, causing the gradient to become zero and making parameter updates difficult.

To address this issue, Someki et al~\cite{context_aware} proposed using the Straight-through Gumbel-softmax Estimator (SGSE)~\cite{gumbel_softmax},
which enables the pruning mask to remain binary during training,
thereby ensuring consistency between training and inference computations. 
Specifically, letting $t(\cdot)$ denote the SGSE function, the output $\tilde{y}$ is computed as 
$\tilde{y} = f(x; \theta \odot t(g(\cdot)))$.

Since $t(\cdot)$ is based on the softmax function, it cannot compute a probability for a single value like the sigmoid function.
To overcome this, the output of $g(\cdot)$ is formulated as a two-class classification problem, and the probability for one class is used as the pruning mask.
During inference, a similar computation is performed using the softmax function, where a threshold is applied to the probability of the class representing the pruning mask to determine the final binary pruning mask.
This approach ensures that the pruning decision follows the same computational flow in both training and inference, maintaining consistency and stability in the pruning process.

\subsection{Global Gate Predictor.}
\label{para:global_gate_predictor}

In globalGP~\cite{context_aware}, the parameter set $\theta$ encompasses the modules within both the encoder and decoder, with the pruning probabilities for all modules computed simultaneously.
Specifically, the set of encoder modules is defined as $\mathcal{M}_{\text{enc}} = \{\text{self-attn}^i, \text{cgMLP}^i, ...\}$, comprising the feed forward network (FFN), self-attention (self-attn) and MLP with convolutional gating (cgMLP)~\cite{cgmlp} in the $i$-th E-Branchformer~\cite{e_branchformer} layers.
Here, the superscript represents the layer index.
Similarly, the decoder module set is defined as $\mathcal{M}_{\text{dec}} = \{\text{self-attn}^i, \text{src-attn}^i, ...\}$, which includes FFN, self-attn, and source attention (src-attn) mechanisms.
Then the outputs $x_{\text{enc}}^i$ and $x_{\text{dec}}^i$ from one of the modules in $i$-th layer $\texttt{m}_{\text{enc}}^i \in \mathcal{M}_{\text{enc}}$ and $\texttt{m}_{\text{dec}}^i \in \mathcal{M}_{\text{dec}}$ of encoder and decoder networks, respectively, are computed as follows:
\begin{align}
    \label{eq:global_mask}
    z_{\text{enc}}^i &= t(g(x,\;C)),\quad &z_{\text{dec}}^i &= t(g(x,\;C)), \\ 
    x_{\text{enc}}^i &= \texttt{m}^i_{\text{enc}}(x_{\text{enc}}^{i-1}, z_{\text{enc}}^i), 
&x_{\text{dec}}^i &= \texttt{m}^i_{\text{dec}}(x_{\text{dec}}^{i-1}, z_{\text{dec}}^i).
\end{align}

Here, $z_{\text{enc}}^i$ and $z_{\text{dec}}^i$ serve as pruning masks for the modules in the $i$-th layer of the encoder and decoder, respectively.
The function $g(x)$ outputs the pruning probabilities for all modules simultaneously.

%% file: blocks/algorithm1.tex
\begin{algorithm}[!t]
\caption{Local Gate Predictor for layer $i$}
\label{alg:local_gp}
\begin{algorithmic}
\Require $C_{\text{key}} \in \mathbb{R}^{T \times D^C \times D},\text{ }C_{\text{value}} \in \mathbb{R}^{T \times D^C \times D}$
\Require $x^i \in \mathbb{R}^{T \times 1 \times D}$  \Comment \small $x_i$ \text{is used as a query.}
\State $\text{gates} \gets []$
\State $a^i \gets \text{Softmax}(x^i \times C_{\text{key}}^\top, \text{ dim}=-1)$ \Comment $\mathbb{R}^{T \times 1 \times D^C}$
\State $a^i \gets a^i \times C_{\text{value}} + x^i$ \Comment $\mathbb{R}^{T \times 1 \times D}$
\For{$n = 1$ \textbf{to} N}
    \State $p_n \gets t(\text{Linear}_n(a^i), \text{ dim}=-1)$ \Comment $\mathbb{R}^{T \times 1 \times 2}$
    \State $\text{gates}.\text{append}(p_n[0])$
\EndFor
\State $C_{\text{key}} \gets \texttt{concat}((C_{\text{key}}, \text{ } x^i), \text{ dim}=1)$
\State $C_{\text{value}} \gets \texttt{concat}((C_{\text{value}}, \text{ Linear}(x^i)), \text{ dim}=1)$
\State \textbf {return} $\text{gates},\text{ } C_{\text{key}},\text{ }C_{\text{value}}$

\end{algorithmic}
\end{algorithm}

%% file: sections/local_gate_predictor.tex
\section{Proposed Method}
\label{sec:methods}

\subsection{Local Gate Predictor}
\label{ssec:local_gate_predictor}

In localGP, the pruning mask $z$ is computed separately for each layer.
Consequently, Equation~\ref{eq:global_mask} is modified as follows:
\begin{align}
\label{eq:reformulation_loss}
    z_{\text{enc}}^i &= t(g(x_{\textcolor{red}{\text{enc}}}^{\textcolor{red}{i-1}}, \textcolor{red}{\hat{C}})),\quad &z_{\text{dec}}^i &= t(g(x_{\textcolor{red}{\text{dec}}}^{\textcolor{red}{i-1}}, \textcolor{red}{\hat{C}})).
\end{align}
In globalGP, the context $C$ represents discrete identifiers such as language ID or task ID, often implemented as special tokens.
However, in this work, we introduce $\hat{C}$ to represent a richer set of contextual information, such as speaker embeddings and acoustic event information.
LocalGP generates a separate pruning mask for each layer, enabling adaptive pruning customized to the characteristics of individual layers.
To enhance contextual awareness, we modify the input by incorporating $x^i$, enabling each layer to access and leverage information accumulated from previous layers.

Algorithm~\ref{alg:local_gp} details the process,
where $D$ is the dimensionality of the context,
$T$ is the sequence length,
$D^C$ is the number of contexts used,
and $N$ is the number of modules inside the $i$-th layer.
We use $C$ as $C_{\text{key}}$ and apply a linear transformation to $C_{\text{value}}$, which forms key-value pairs.
The input feature $x^i$ to the $i$-th layer is used as the query.
Regarding $D^C$, for example, when using speaker embeddings and acoustic event embeddings, we set $D^C = 2$.
Cross-attention is computed with a residual connection, and $z$ is derived from $t$. A linear layer is applied to compute the logits.
Finally, the results are added to $C_{\text{key}}$ and $C_{\text{value}}$.

\input{blocks/table_1_colored4}
\subsection{Temporal pruning vs. utterance-wise pruning}
\label{ssec:frame_vs_utt}

Let $\texttt{m}_n \in \mathcal{M}_{\text{enc}} \cup \mathcal{M}_{\text{dec}}$ be a module, $\tilde{y}_n$, $x_n$, and $z_n$ be the input, output, and the pruning mask of the module $\texttt{m}_n$, respectively.
Then, $\tilde{y}_n$ in temporal pruning and utterance-wise pruning becomes:
\small
\label{eq:frame_vs_utt}
\begin{align}
\tilde{y}_n =
\begin{cases}
  \texttt{pad}(\texttt{m}(\texttt{h}(x_n,\;z_n)),\;z_n), & \text{if pruning by frame,} \\ 
  x_n * 0, & \text{if pruning by utt, } z_n = 0, \\
  \texttt{m}(x_n), & \text{if pruning by utt, } z_n = 1.
\end{cases}
\end{align}
\normalsize
The \verb|h| function selects non-skipped frames, while the \verb|pad| function zero-pads the output tensor based on the pruning mask.
If the number of indices selected by \verb|h| is smaller than the convolution kernel size in cgMLP, computation becomes infeasible.
To address this, the cgMLP module is always computed for all frames in our work.

%% file: blocks/table_1_colored4.tex
\begin{table*}[!htp]\centering
\caption{WER and BLEU scores for ASR and ST in German (de), French (fr), and Italian (it) with utterance-wise pruning (globalGP) and temporal pruning (localGP) strategies on OWSM-v3.1.
The \texttt{Context} column denotes the additional context used for pruning.
\texttt{Enc+Dec} indicates that pruning is applied to both encoder and decoder modules, while \texttt{front} refers to subsampled speech features from the frontend.
\texttt{spk} and \texttt{event} indicate that pruning is guided by speaker embeddings and acoustic event information, respectively.
}
\label{tab:table_1}
\footnotesize
\resizebox{\linewidth}{!}{
\begin{tabular}{clclcccccccccccccc}\toprule
\multirow{2}{*}{No.} &\multirow{2}{*}{\shortstack{Pruned\\Module}} &\multirow{2}{*}{Is localGP?} &\multirow{2}{*}{Context} &\multicolumn{4}{c}{ASR-WER ($\downarrow$)} & &\multicolumn{7}{c}{ST ($\uparrow$)} &\multirow{2}{*}{GFLOPs (Enc)} \\\cmidrule{5-8}\cmidrule{10-16}
& & & &de &fr &it &Average & &de-fr &de-it &fr-de &fr-it &it-de &it-fr &Average & \\\midrule
\cellcolor[HTML]{ecf9fa}1 &\cellcolor[HTML]{ecf9fa}- &\cellcolor[HTML]{ecf9fa} &\cellcolor[HTML]{ecf9fa}full fine-tuning (baseline) &\cellcolor[HTML]{ecf9fa}13.6 &\cellcolor[HTML]{ecf9fa}11.0 &\cellcolor[HTML]{ecf9fa}13.2 &\cellcolor[HTML]{ecf9fa}12.6 &\cellcolor[HTML]{ecf9fa} &\cellcolor[HTML]{ecf9fa}8.4 &\cellcolor[HTML]{ecf9fa}6.4 &\cellcolor[HTML]{ecf9fa}11.2 &\cellcolor[HTML]{ecf9fa}13.0 &\cellcolor[HTML]{ecf9fa}11.8 &\cellcolor[HTML]{ecf9fa}13.5 &\cellcolor[HTML]{ecf9fa}10.7 &\cellcolor[HTML]{ecf9fa}568.5 \\
\midrule \midrule
\multicolumn{17}{c}{Utterance-wise pruning (GlobalGP)} \\ \midrule
\cellcolor[HTML]{fefbf2}2 &\cellcolor[HTML]{fefbf2}Encoder &\cellcolor[HTML]{fefbf2} &\cellcolor[HTML]{fefbf2}front (baseline) &\cellcolor[HTML]{fefbf2}15.0 &\cellcolor[HTML]{fefbf2}11.9 &\cellcolor[HTML]{fefbf2}15.2 &\cellcolor[HTML]{fefbf2}14.0 &\cellcolor[HTML]{fefbf2} &\cellcolor[HTML]{fefbf2}5.8 &\cellcolor[HTML]{fefbf2}6.0 &\cellcolor[HTML]{fefbf2}8.6 &\cellcolor[HTML]{fefbf2}12.3 &\cellcolor[HTML]{fefbf2}6.5 &\cellcolor[HTML]{fefbf2}12.6 &\cellcolor[HTML]{fefbf2}8.6 &\cellcolor[HTML]{fefbf2}452.5 \\
\cellcolor[HTML]{fefbf2}3 &\cellcolor[HTML]{fefbf2}Decoder &\cellcolor[HTML]{fefbf2} &\cellcolor[HTML]{fefbf2}front (baseline) &\cellcolor[HTML]{fefbf2}15.2 &\cellcolor[HTML]{fefbf2}11.9 &\cellcolor[HTML]{fefbf2}12.9 &\cellcolor[HTML]{fefbf2}13.3 &\cellcolor[HTML]{fefbf2} &\cellcolor[HTML]{fefbf2}9.8 &\cellcolor[HTML]{fefbf2}8.4 &\cellcolor[HTML]{fefbf2}12.2 &\cellcolor[HTML]{fefbf2}13.1 &\cellcolor[HTML]{fefbf2}11.3 &\cellcolor[HTML]{fefbf2}13.4 &\cellcolor[HTML]{fefbf2}11.4 &\cellcolor[HTML]{fefbf2} - \\
\cellcolor[HTML]{fefbf2}4 &\cellcolor[HTML]{fefbf2}Enc+Dec &\cellcolor[HTML]{fefbf2} &\cellcolor[HTML]{fefbf2}front (baseline) &\cellcolor[HTML]{fefbf2}15.0 &\cellcolor[HTML]{fefbf2}12.6 &\cellcolor[HTML]{fefbf2}14.6 &\cellcolor[HTML]{fefbf2}14.1 &\cellcolor[HTML]{fefbf2} &\cellcolor[HTML]{fefbf2}7.3 &\cellcolor[HTML]{fefbf2}4.7 &\cellcolor[HTML]{fefbf2}9.9 &\cellcolor[HTML]{fefbf2}9.5 &\cellcolor[HTML]{fefbf2}8.7 &\cellcolor[HTML]{fefbf2}11.5 &\cellcolor[HTML]{fefbf2}8.6 &\cellcolor[HTML]{fefbf2} - \\
\midrule \midrule
\multicolumn{17}{c}{Temporal pruning (LocalGP)} \\ \midrule
\cellcolor[HTML]{fbfff7}5 &\cellcolor[HTML]{fbfff7}Encoder &\cellcolor[HTML]{fbfff7}\checkmark &\cellcolor[HTML]{fbfff7}front &\cellcolor[HTML]{fbfff7}14.6 &\cellcolor[HTML]{fbfff7}11.4 &\cellcolor[HTML]{fbfff7}13.0 &\cellcolor[HTML]{fbfff7}13.0 &\cellcolor[HTML]{fbfff7} &\cellcolor[HTML]{fbfff7}10.4 &\cellcolor[HTML]{fbfff7}7.8 &\cellcolor[HTML]{fbfff7}13.0 &\cellcolor[HTML]{fbfff7}14.6 &\cellcolor[HTML]{fbfff7}11.2 &\cellcolor[HTML]{fbfff7}15.3 &\cellcolor[HTML]{fbfff7}12.0 &\cellcolor[HTML]{fbfff7}541.6 \\
\cellcolor[HTML]{fbfff7}6 & \cellcolor[HTML]{fbfff7}Encoder&\cellcolor[HTML]{fbfff7}\checkmark &\cellcolor[HTML]{fbfff7}+ spk &\cellcolor[HTML]{fbfff7}14.5 &\cellcolor[HTML]{fbfff7}\textbf{11.1} &\cellcolor[HTML]{fbfff7}\textbf{12.7} &\cellcolor[HTML]{fbfff7}12.8 &\cellcolor[HTML]{fbfff7}\textbf{} &\cellcolor[HTML]{fbfff7}\textbf{12.0} &\cellcolor[HTML]{fbfff7}\textbf{9.2} &\cellcolor[HTML]{fbfff7}\textbf{14.4} &\cellcolor[HTML]{fbfff7}\textbf{15.6} &\cellcolor[HTML]{fbfff7}\textbf{12.7} &\cellcolor[HTML]{fbfff7}\textbf{16.9} &\cellcolor[HTML]{fbfff7}\textbf{13.5} &\cellcolor[HTML]{fbfff7}511.8 \\
\cellcolor[HTML]{fbfff7}7 & \cellcolor[HTML]{fbfff7}Encoder&\cellcolor[HTML]{fbfff7}\checkmark &\cellcolor[HTML]{fbfff7}+ event &\cellcolor[HTML]{fbfff7}\textbf{14.1} &\cellcolor[HTML]{fbfff7}\textbf{11.1} &\cellcolor[HTML]{fbfff7}13.0 &\cellcolor[HTML]{fbfff7} \textbf{12.7} &\cellcolor[HTML]{fbfff7} &\cellcolor[HTML]{fbfff7}11.4 &\cellcolor[HTML]{fbfff7}8.2 &\cellcolor[HTML]{fbfff7}13.5 &\cellcolor[HTML]{fbfff7}15.1 &\cellcolor[HTML]{fbfff7}12.0 &\cellcolor[HTML]{fbfff7}16.3 &\cellcolor[HTML]{fbfff7}12.8 &\cellcolor[HTML]{fbfff7}510.1 \\
\cellcolor[HTML]{fbfff7}8 & \cellcolor[HTML]{fbfff7}Encoder&\cellcolor[HTML]{fbfff7}\checkmark &\cellcolor[HTML]{fbfff7}+ spk + event &\cellcolor[HTML]{fbfff7}15.2 &\cellcolor[HTML]{fbfff7}12.4 &\cellcolor[HTML]{fbfff7}14.5 &\cellcolor[HTML]{fbfff7}14.0 &\cellcolor[HTML]{fbfff7} &\cellcolor[HTML]{fbfff7}10.6 &\cellcolor[HTML]{fbfff7}8.1 &\cellcolor[HTML]{fbfff7}13.6 &\cellcolor[HTML]{fbfff7}15.0 &\cellcolor[HTML]{fbfff7}12.0 &\cellcolor[HTML]{fbfff7}16.3 &\cellcolor[HTML]{fbfff7}12.6 &\cellcolor[HTML]{fbfff7}497.0 \\
\cellcolor[HTML]{fbfff7}9 &\cellcolor[HTML]{fbfff7}Decoder &\cellcolor[HTML]{fbfff7}\checkmark &\cellcolor[HTML]{fbfff7}lang2vec &\cellcolor[HTML]{fbfff7}15.0 &\cellcolor[HTML]{fbfff7}11.2 &\cellcolor[HTML]{fbfff7}\textbf{12.7} &\cellcolor[HTML]{fbfff7}13.0 &\cellcolor[HTML]{fbfff7} &\cellcolor[HTML]{fbfff7}11.2 &\cellcolor[HTML]{fbfff7}8.4 &\cellcolor[HTML]{fbfff7}13.7 &\cellcolor[HTML]{fbfff7}15.2 &\cellcolor[HTML]{fbfff7}12.3 &\cellcolor[HTML]{fbfff7}16.4 &\cellcolor[HTML]{fbfff7}12.9 &\cellcolor[HTML]{fbfff7} - \\
\cellcolor[HTML]{fbfff7}10 &\cellcolor[HTML]{fbfff7}Enc+Dec &\cellcolor[HTML]{fbfff7}\checkmark &\cellcolor[HTML]{fbfff7}front + event + lang2vec &\cellcolor[HTML]{fbfff7}14.5 &\cellcolor[HTML]{fbfff7}11.0 &\cellcolor[HTML]{fbfff7}13.1 &\cellcolor[HTML]{fbfff7}12.9 &\cellcolor[HTML]{fbfff7} &\cellcolor[HTML]{fbfff7}10.5 &\cellcolor[HTML]{fbfff7}8.0 &\cellcolor[HTML]{fbfff7}13.1 &\cellcolor[HTML]{fbfff7}14.9 &\cellcolor[HTML]{fbfff7}11.9 &\cellcolor[HTML]{fbfff7}16.0 &\cellcolor[HTML]{fbfff7}12.4 &\cellcolor[HTML]{fbfff7} - \\
\bottomrule
\end{tabular}
}
\end{table*}
\vspace{-10pt}

%% file: sections/experiments.tex
\section{Experiments}


\subsection{Models.}
\label{ssec:exp_models}
In this study, we employed version 3.1 of OWSM model~\cite{owsm_v31}, an open-source alternative to OpenAI’s Whisper~\cite{radford2023robust}. 
We chose OWSM-v3.1 for its openness and reproducibility; unlike Whisper, it is trained entirely on publicly available data.
This allows us to ensure that our evaluation set was not included in pretraining, enabling a fair assessment of pruning effectiveness.
To the best of our knowledge, no other public encoder-decoder model with attention-CTC is available for such comparison.


\subsection{Experimental Setups.}
\label{ssec:exp_dataset_and_tasks}
We used version 1.1 of the Europarl-ST~\cite{europarl_st} corpus to evaluate our method and the baseline.
We selected French, German, and Italian among the 9 languages in this corpus to ensure a language-balanced training data.  
Each language contains approximately 20 hours of speech data.
With this dataset, we fine-tuned all models with a pruning objective across ASR and speech translation (ST) tasks.
We used a batch size of 4, Adam optimizer, Warmup LR Scheduler with learning rate of $1e^{-5}$ and 6000 warmup steps.
In all experiments, we followed~\cite{context_aware} and used OWSM-v3.1 as the backbone model with a target sparsity of 30\%, meaning that 70\% of the model’s modules or frames remained active during training and inference.

To better capture audio information for pruning on the encoder, and language information for decoder, we utilized the 
SOTA models for each context:
ECAPA\_TDNN~\cite{desplanques20_interspeech} for speaker embedding;
BEATs~\cite{chen2023beats} for acoustic events information;
and URIEL lang2vec~\cite{littell2017uriel} for language information.
We specifically use BEATs' second layer because we want to keep the context extraction module small.
When the dimensionalities of speaker embeddings and acoustic event embeddings differ from $D^C$, we use a linear layer to align them.
Similarly, if the sequence lengths vary, we either duplicate the final frame or trim the sequence to match the length of the audio.


We evaluate ASR performance with Word Error Rate (WER) and ST using BLEU scores, while measuring encoder GFLOPs to assess the impact of pruning and additional context.
Since OWSM uses autoregressive decoding with fixed 30-second input speech, external factors like hypothesis count and end-detection mechanisms are controlled.
To isolate the effect of pruning strategies and additional context, we specifically measure the encoder GFLOPs rather than the entire model.
This allows us to clearly visualize the computational differences introduced by temporal pruning, utterance-wise pruning, and contextual inputs.
A beam size of 5 was used for evaluation.

\subsection{Results}
\label{para:exp_results}
In Table~\ref{tab:table_1}, we compare row 2 and 5 to evaluate the effect of temporal pruning.
Temporal pruning with LocalGP leads to a significant performance improvement in both ASR and ST, achieving an average relative improvement of 39.6\% in BLEU score.
Additionally, ASR exhibits an average performance gain of 7.3\% in WER.
In terms of GFLOPs reduction, temporal pruning does not achieve the same level of reduction as globalGP by comparing the row 2 and 5 to 8.
This is likely because, in utterance-wise pruning, the computations for pruned modules are entirely skipped, whereas in temporal pruning, all modules are still computed.
Nevertheless, simply applying our proposed method effectively reduces 26.9 GFLOPs from the top-line OWSM-v3.1 model while achieving a 25.7\% relative improvement in BLEU, demonstrating its efficiency in pruning.
We also measured the averaged encoder-side wall-clock time: the model with speaker embedding (row 6) took 0.124s, slightly higher than globalGP's 0.111s (row 2).


Next, we examine the characteristics of different acoustic features by comparing rows 5 through 8.
Overall, the addition of speaker embeddings in row 6 led to a significant performance improvement.
Compared to the baseline in row 2, ASR achieved an average relative WER reduction of 9.0\%, while translation tasks exhibited an average relative BLEU score improvement of 56.0\%.
A similar trend was observed with acoustic event information (row 7), where ST achieved an average relative BLEU score gain of 47.7\%.
These results suggest that rather than directly using subsampled speech features, leveraging a pre-trained model to extract rich contextual information enables more effective selection of frames for pruning.
For ST, incorporating speaker embeddings significantly outperformed the fully fine-tuned OWSM-v3.1 model, achieving a relative BLEU score improvement of 28.6\%.
Speaker embeddings and acoustic event features also showed a similar trend in GFLOPs reduction, achieving reductions of 56.7 GFLOPs and 58.4 GFLOPs, respectively.
These findings demonstrate that our proposed method improves both inference flops and the performance of the OWSM-v3.1 encoder.

\input{blocks/figure_1}


Comparing decoder results in Table~\ref{tab:table_1}, rows 4 and 10, we find that temporal pruning also improves performance.
However, due to batch-wise beam search, different pruning masks were applied to different beams.
As a result, most decoder modules were still computed, negating expected inference speed gains.
A tailored implementation could theoretically accelerate decoder-side pruning as well.

\subsection{Temporal Pruning Pattern Analysis for Encoder}
\label{ssec:pruning_pattern_encoder}
We visualized how temporal pruning is performed when using the speaker embedding model as a context.
Figure~\ref{fig:logmel_att} illustrates an example from the test set, showing the log-mel spectrogram and the pruning mask of the self-attention module.
The figure reveals that frames are actively utilized during speech segments, while relatively fewer computations are allocated to silence regions.
Notably, the first layer attends to almost all frames, whereas the deeper layers exhibit more pronounced pruning patterns, suggesting that the initial layer has acquired a VAD-like function.
However, unlike conventional VAD systems, certain layers, particularly in the middle to later stages, exhibit non-negligible attention to silence frames, suggesting that these layers encode silence-related features.
This indicates that the model selectively activates frames in layers responsible for capturing silence-related information, achieving optimal temporal pruning at both the layer and module levels.

As discussed in Section~\ref{para:exp_results}, temporal pruning on the encoder side appears to have a similar effect to VAD system.
This suggests that adding speaker embeddings may have allowed the model to respond more sensitively to the speaker’s voice activity.
However, when combining these two feature types in row 9, no significant performance improvement was observed.
Although GFLOPs were lower than those of the two individual models, considering the additional computation required for two pre-trained models, this approach cannot be regarded as an effective reduction in computational cost.
Furthermore, performance was comparable to or even worse than that of row 8.
These findings indicate that multiple audio context features does not necessarily lead to performance improvement, and it is sufficient to select a single model that with the best results.


\input{blocks/figure_2}

\input{blocks/table_stats}
\subsection{Analysis on Decoder result}
\label{ssec:exp_frame_decoder}
We also focused on the decoder side to examine how pruning is performed for each token. As shown in Figure~\ref{fig:token_space}, we found that the usage rate of source-attention differs depending on whether a token begins with a space, meaning that the token is the start of a new word.
To verify this, we analyzed the relationship between the sparsity ratio of source-attention and the presence of spaces in the tokens generated in the German ASR test set. 
Among all output tokens, 27,032 contained space, while 30,667 did not.
We conducted statistical tests to assess this difference, as summarized in Table~\ref{tab:src_att_stats}. The results indicate a significant difference in sparsity ratio between tokens with and without spaces. 
Specifically, tokens starting a new word require more attention to the encoder output, indicating that the model relies more on audio information for these tokens.
This suggests that the model dynamically adjusts its reliance on specific modules based on token traits, which could have implications for optimizing decoding efficiency.
Furthermore, the tendency to suppress source-attention of early decoder layers is consistent with (Someki et al., 2025).
We hypothesize that localGP amplifies this effect by utilizing the required amount of source attention for each tokens, reducing encoder-induced noise during decoding and acting as a regularizer, particularly improving ST.

%% file: blocks/figure_1.tex
\begin{figure}[!t]
\centering
\includegraphics[width=\linewidth]{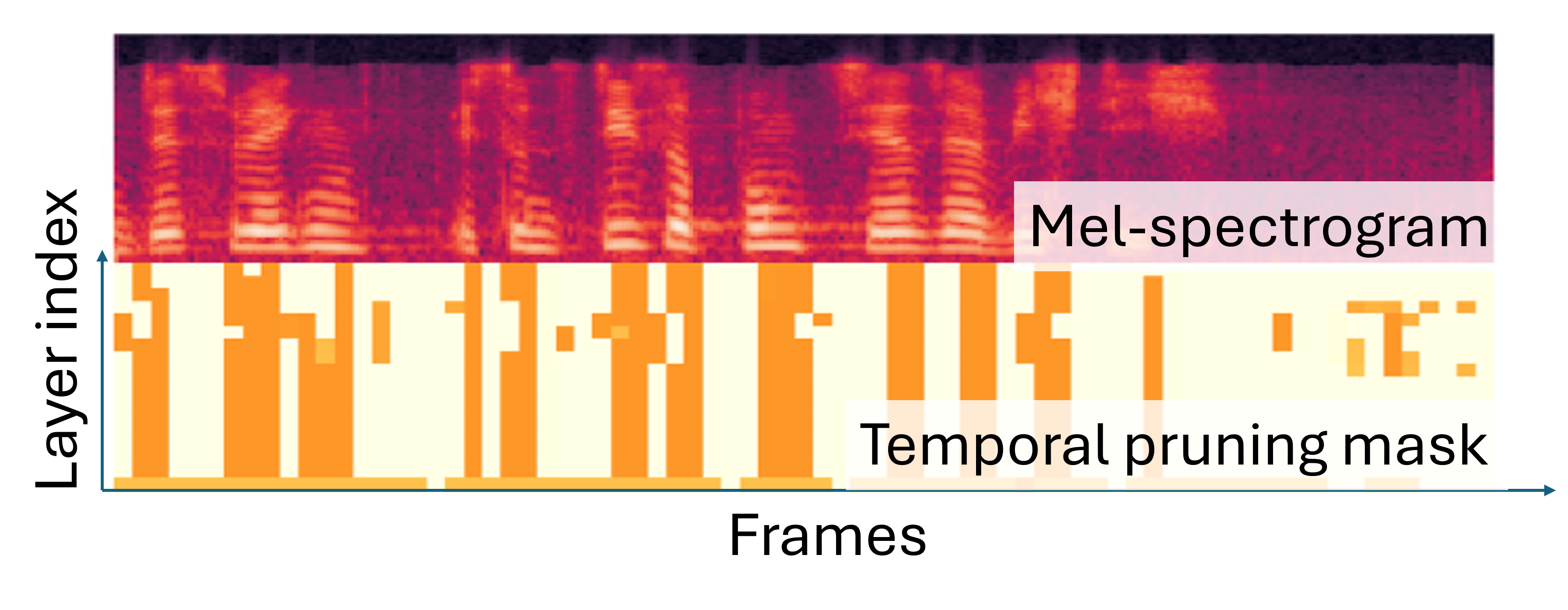}
\caption{
Log-Mel spectrogram (top) and temporal pruning mask for self-attention modules (bottom).
The y-axis of the lower plot represents the layers, with the initial layer at the bottom.
The x-axis of both plots represents the time scale.
The orange regions in the lower plot indicate activated self-attention modules, while the white regions represent pruned modules.
}
\label{fig:logmel_att}
\end{figure}

%% file: blocks/figure_2.tex
\begin{figure}[!t]
    \centering
    \begin{subfigure}[b]{0.48\linewidth}
        \centering
        \includegraphics[width=\linewidth]{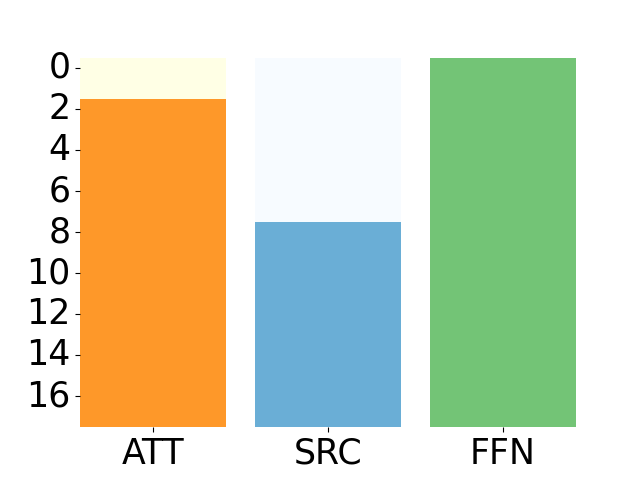}
        \caption{Token: \texttt{[Space]wollen}}
        \label{fig:wollen}
    \end{subfigure}
    \hfill
    \begin{subfigure}[b]{0.48\linewidth}
        \centering
        \includegraphics[width=\linewidth]{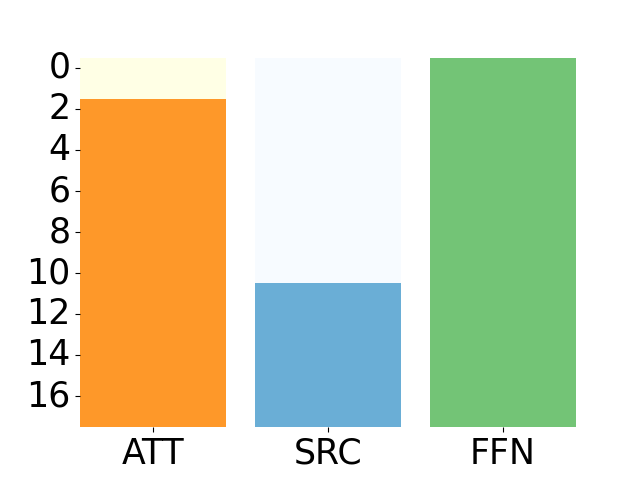}
        \caption{Token: \texttt{struktur}}
        \label{fig:struktur}
    \end{subfigure}
    \caption{Pruning pattern for tokens \texttt{[Space]wollen} and \texttt{struktur}.
The \texttt{ATT}, \texttt{SRC}, and \texttt{FFN} represent self-attention, source-attention, and the feed-forward network inside the decoder, respectively.
The y-axis indicates the layers, with the top representing the first layer.
Colored modules indicate activated modules.
The number of \texttt{SRC} modules is higher when a space precedes a token.}
    \label{fig:token_space}
\end{figure}
\vspace{-5pt}

%% file: blocks/table_stats.tex
\begin{table}[!tp]\centering
\caption{Statistical test results on source-attention usage}
\label{tab:src_att_stats}
\begin{tabular}{lrcc}\toprule
Test & Test statistic & p-value \\\midrule
Mann-Whitney U test & $5.38 \times 10^8$ & $< 0.0001$ \\
Welch’s t-test & 70.9 & $< 0.0001$ \\
\bottomrule
\end{tabular}
\end{table}

%% file: sections/conclusion.tex
\section{Conclusion}
\label{sec: conclusion}
In this study, we proposed localGP, a context-driven dynamic inference optimization method that integrates external context, including speaker embeddings, acoustic events, and linguistic information.
Our experiments demonstrated that combining localGP with temporal pruning and speaker embeddings as additional context reduced computation by 56.7 GFLOPs compared to the original OWSM-v3.1. 
Additionally, we achieved ST performance exceeded that of the fine-tuned baseline OWSM-v3.1, with BLEU score relatively improvements of 25.6\% on average.
Furthermore, our analysis of the pruning masks showed that the first layer of the OWSM encoder acquired a VAD-like function, confirming that our method dynamically optimizes model computation based on input speech characteristics.
